\documentclass[10pt,a4paper]{book}
\usepackage{arxiv}
\begin{document}
\newcounter{ctr}
\setcounter{ctr}{\thepage}
\addtocounter{ctr}{5}

\talktitle{
Testing for the Pioneer anomaly on a Pluto exploration mission}
\talkauthors{Andreas Rathke \structure{a}} 

\begin{center}
\authorstucture[a]{ESA Advanced Concepts Team (DG-X), ESTEC,
Noordwijk, The Netherlands}
\end{center}

\shorttitle{Testing for the Pioneer anomaly on a Pluto exploration
  mission} 

\firstauthor{A. Rathke}

\vspace*{-1cm}
\begin{abstract}
The Doppler-tracking data of the Pioneer 10 and 11 spacecraft show an
unmodelled constant acceleration in the direction of the inner Solar
System. 
An overview of the phenomenon,
commonly dubbed the Pioneer anomaly, is given and the possibility for
an experimental test of the anomaly as a secondary goal of an upcoming
space mission is discussed using a putative Pluto orbiter probe as a
paradigm.
\end{abstract}

\vspace*{-1cm}\section{Introduction}

The orbit reconstruction from the Doppler tracking data of the
hyperbolic trajectory away from the Sun of the Pioneer~10 and~11
probes shows an anomalous deceleration of both spacecraft of the order
of $10^{-9}\,{\rm m/s^2}$.  Even before the Jupiter swing-by, an
unmodelled acceleration of that order \cite{Null} had been noticed
for Pioneer 10. It had however been attributed to gas leaks and a
mismodelling of the solar radiation force. Such patterns of
explanation became unsatisfactory for the post swing-by hyperbolic arc due
to the decrease of the solar radiation pressure inversely proportional
to the square of the distance from the Sun and the quiet state of the
spacecraft, which should reduce any gas leaks.  The anomaly on both
probes has been subject to three independent analyses
\cite{Anderson:1998jd,Markwardt:2002ma}. The result of all
investigations is that an anomalous Doppler blueshift is present in
the data from both craft of approximately $1.1 \times 10^{-8}\,{\rm
Hz/s}$ corresponding to an apparent deceleration of the spacecraft of
approximately $8 \times 10^{-10}\,{\rm m/s}^2$.  From Doppler data
alone, it is not possible to distinguish between an anomalous
frequency shift of the radio signal or a real deceleration of the
spacecraft.

\begin{figure}[t]  
\begin{minipage}[b]{0.35\linewidth}
\centering
\includegraphics[height=5cm]{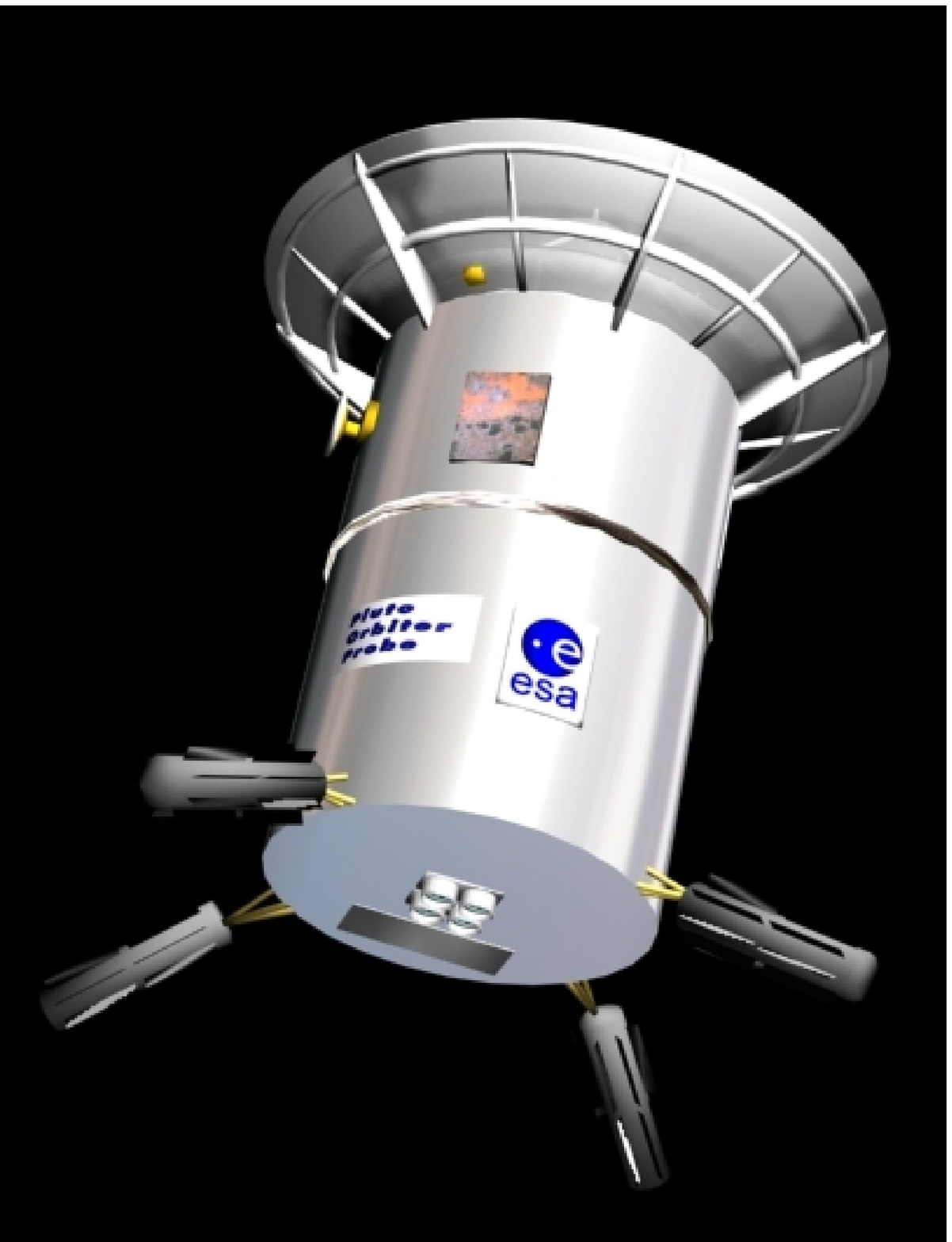}
\caption{View of POP.}    \label{fig1}
\end{minipage}
\begin{minipage}[b]{0.65\linewidth}
\centering
\includegraphics[height=5.5cm]{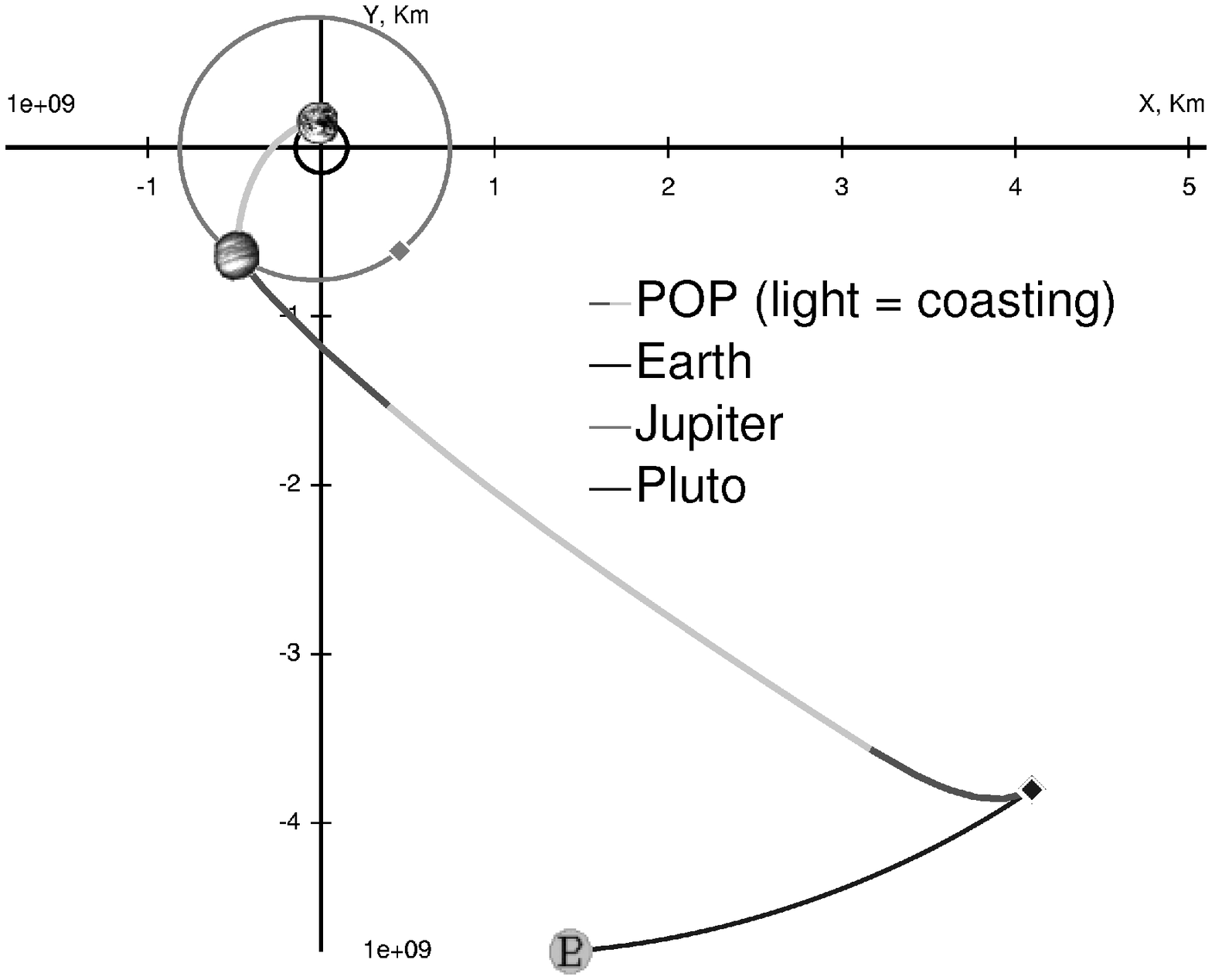}
\vspace*{-5mm}
\caption{The trajectory of POP.}    \label{fig2}
\end{minipage}\vspace*{-5mm}
\end{figure}

The principle investigators of the anomaly have conducted a thorough
investigation of possible biases and concluded that no conventional
effect is likely to have caused the anomaly
\cite{Anderson:1998jd}. 
Meanwhile, there exists an ample body of literature discussing various aspects
of possible systematic effects, 
without definitive conclusion.

Although the Pioneer anomaly (PA) is an effect at the edge of what is
detectable with radiometric tracking of a deep-space probe, it is huge
in physical terms: The anomaly exceeds the general-relativistic
corrections to Newtonian motion by
five orders of magnitude (at 50\,AU). 
A gravitational origin of the deceleration of the Pioneer probes is
however hard to imagine, since no corresponding anomaly is seen in the
orbits of Uranus and Neptune. Hence, a gravitational anomalous
deceleration would indicate a violation of the weak equivalence
principle.

Considering the efforts that have been undertaken to find a
conventional explanation of the PA, it seems likely that
only an experiment will finally be able to determine the nature of the
effect. A mission to perform an investigation of the PA
has to exceed the navigational accuracy of the Pioneers. In particular,
the systematic errors in the modelling of onboard generated forces
must not exceed a few percent of the Pioneers' anomalous deceleration
and the test should take place in the outer part of the Solar
system so that external disturbances are minimised.  

The spacecraft currently in operation or in design are not
capable 
to fulfil these requirements. Hence, a
PA test will have to be performed by a mission which takes
into account the experiment already in an early stage of its design process.
We discuss in the following how the test requirements can be
achieved on a mission which has a test of the PA only
as a secondary goal. Considering a non-dedicated mission as a first
choice seems reasonable in view of the high cost of a mission to the
outer Solar system.

\section{The Pluto orbiter probe}

To make our considerations about the implementation of a PA test as
tangible as possible we consider its realisation onboard of a
low-mass, low-thrust mission to Pluto.  A study of such a mission has
been undertaken recently in ESA's Advanced Concepts Team and detailed
results of the system design and trajectory design have been presented
in \cite{Bondo}.

The goal of the Pluto-orbiter-probe (POP) study was to design a
spacecraft that would reach a low circular orbit around Pluto,
1000\,km above its surface, using only hardware that will already be
space-qualified by the launch date.  The capability of entering a
Pluto orbit is provided by a nuclear electric propulsion system.  The
trajectory will incorporate a Jupiter gravity assist.
The envisaged launch
date is in 2016, arrival at Pluto would be 18 years later in 2034.
The spacecraft's wet mass is 837\,kg, and the dry mass 516\,kg. This would
enable a launch with a heavy
launcher with an Earth escape velocity of $10\,{\rm km/s}$.  The
science payload will have a mass of 20\,kg and will feature a 
multi-spectral imaging camera, a near-infrared spectrometer, an X-ray
spectrometer, a bolometer and the provision to use the communication
antenna also as a synthetic-aperture RADAR.

The central part of POP (Fig.~\ref{fig1}) 
is a cylindrical spacecraft bus of 1.85\,m
length and 1.2\,m diameter.  The 2.5\,m diameter Ka-band (32\,GHz) high-gain
antenna is located at one end of the bus.
The other end of the spacecraft bus houses the propulsion system.  It
consists of four radioisotope thermoelectric generators (RTG's) on
short booms inclined $45^\circ$ to the axis of the bus, a toroidal
tank for the Xenon propellant and the four QinetiQ~T5 ion thruster
main engines.
Attitude control is provided by 10 hollow-cathode thrusters, which use the
same power-processing unit and propellant supply as the main engines.

The trajectory of POP (Fig.~\ref{fig2}) 
has a hyperbolic coast phase of 
18\,AU length before the braking burn for the
Pluto orbit capture begins. This coast phase will last 7.4 years.
During this time the spacecraft will transverse a radial distance
between 13.4\,AU and 30.4\,AU from the Sun.
The mass of POP will then be about 760\,kg.

POP is envisaged to employ two different attitude-control modes
during different parts of its journey. When the main engines of POP
are thrusting and during the swing-by the craft will be three-axis
stabilised in order improve pointing accuracy. The same is the case in
Pluto orbit where reorientation requirements demand
three-axis control. During long coasts, however, spin stabilisation
will be employed. This has the practical benefit of enhancing the
lifetime of the momentum wheels and, thus, saves mass. 
The rotational speed will be very low during the coast, ca.\ 0.5\,rpm,
so that attitude acquisition can still be preformed by the star
trackers to be used in 3-axis stabilised mode. Hence, no additional
attitude acquisition hardware will be necessary during the spinning
mode and it can be realised at no additional spacecraft mass.
Spin stabilisation during the coast is a crucial factor to enable a
test of the PA as it reduces the number of attitude control
manoeuvres to one every few weeks and hence minimises onboard generated
accelerations.

\section{Measurement strategy}

With a new mission, the search for a PA-like effect 
will still rely on radio-tracking. 
Nevertheless, considerable advances over 
the precision of the Pioneer data are achievable
with present-day telecommunication hardware.
The major improvement comes from the use of sequential ranging in
addition to Doppler tracking. 
The information from sequential ranging relies on the group
velocity of the signal, whereas the information from Doppler tracking
relies on the phase velocity. Sequential ranging is
hence insensitive to a gravitational effect on the radio signal, which
would be non-dispersive. The usage of both 
methods allows an anomalous blueshift of the radio
signal to be detected which would affect the Doppler signal only.
Another advantage of sequential ranging is that it allows a
considerably more precise orbit determination than Doppler 
tracking. Interestingly, Delta differential one-way
ranging 
does not add to the performance of a
PA test \cite{nonded}. 

For the POP trajectory, the tracking will be precise enough to
distinguish between a ``drag force'' in the direction opposite to the
velocity vector, and a force pointing towards the centre of the Solar
system. However, despite the high accuracy of tracking techniques,
a discrimination between an Earth and Sun pointing deceleration does
not seem possible with POP.  The distinction would be made by
the search for an annual modulation of the Earth-pointing
component of the acceleration, revealing that the force on the
spacecraft originates from the Sun. For the POP trajectory, the
modulation of a Sun-pointing anomaly would be $\lesssim
0.3\%$. Whereas this is still detectable, the modulation is too low to
show up in the background noise caused by onboard generated systematics (see
next section). Nevertheless the most plausible origins of an anomaly can
be discriminated by the combination of Doppler-tracking and
sequential ranging because an Earth-pointing anomaly should always be
caused by a blueshift and not by an acceleration of the spacecraft.

\section{Overcoming systematics}


For a new mission the magnitude of the anomaly will in general not
coincide with the value from the Pioneer probes but will most likely 
be influenced by the spacecraft design.  
Most importantly, the magnitude of the putative anomaly depends on the 
mass of the satellite for a real acceleration.

For a conventional force, the anomalous acceleration will be inversely
 proportional to the mass of the satellite, in accordance with
 Newton's second principle.  For POP, which is roughly three times as
 heavy as the Pioneer probes during the coast, this would reduce a
 putative anomaly to $3 \times 10^{-10}\,{\rm m/s^2}$.  A mass
 dependence should be present for a gravitational force as well,
 because an explanation in terms of modified gravitation requires a
 violation of the weak equivalence principle. 
If the PA is
 a blueshift of light, the spacecraft mass will not influence the
 magnitude of the anomaly.


Lacking a conclusive theoretical model of the anomaly, no firm prediction
for the magnitude of the anomaly, that one should expect, can be given. 
Thus the spacecraft design has to 
reduce acceleration systematics as far as possible or
provide means to precisely determine the systematics. 
With this goal in mind, we review the major potential sources of systematics
and how they are controlled on POP. 

Fuel leakage is much easier to reduce for an electric
propulsion system than a chemical one. The reason is
that the propellant
for all engines, both main and attitude control, is taken from the same tank
through the same pressure regulator. 
Low leakage rates can easily (and for a moderate mass budget) 
be achieved by stacking several regulators in a row.
In this way the maximal acceleration of POP due to fuel leakage can be
reduced to $\sim 0.1\%$ of the PA.

Whereas electric propulsion systems have a considerable advantage
concerning fuel leakage, they have the major drawback of requiring high
electrical power. For POP, this results in 
17,000\,W of heat, which has to be
dissipated by the radiator fins on the RTG's. Reflection of RTG
thermal radiation by the
spacecraft bus and antenna will generate a deceleration of $8.5
\times 10^{-10}\, {\rm m/s^2}$, the magnitude of the
PA. This force can, however, be discriminated from other
effects because it will decay with the half-life of the Pu in the RTG's
by an amount of 6\% during the measurement coast.  

The radiation force by the 55\,W antenna beam would lead to an
acceleration of $2.5 \times 10^{-10}\,{\rm m/s^2}$.  This force can be
controlled by changing the transmission power during the coast and
measuring the change in the acceleration of the craft. The reduction
in data transmission rate, which accompanies a reduction of the power
is not a problem as only a small amount of housekeeping data needs to be sent
during the cruise.

The Solar radiation force 
is less important 
on POP than 
it was 
on the Pioneers 
%
and
reduces to 1\,\% of the PA at the end of the coast.

Overall the systematic accelerations of POP can be controlled or
modelled to
$10^{-11}\,{\rm m/s^2}$. With this level of systematics it
becomes possible to unambiguously detect 
an anomalous force or blueshift 
of PA magnitude.

\section{Summary and conclusions}

With the current status of our knowledge it would be premature to
consider the PA as a manifestation of a new
physics. Rather, an incorrectly modelled conventional force seems the
most likely origin for the anomaly.
Most lines of ``explanations'' of the PA in terms of
``new physics'' are not stringent, e.\,g.\ it is not at all
clear how to circumvent the constraints from planet orbits if a
real force is present. 
Even without a satisfactory model of the anomaly at hand, we
found that a test for all currently discussed causes of the PA is possible.
The test can be
incorporated in a planetary exploration mission to Pluto at practically
no cost in launch mass, if the objective of the PA test
is taken into account during the design of the spacecraft right from the
beginning.  
Other non-dedicated options for a PA test
will be discussed elsewhere \cite{nonded}.

\paragraph*{Acknowledgements}

The author is grateful to Roger Walker for helpful comments on
the manuscript. This work has benefited from discussions
with Torsten Bondo and Dario Izzo.

\end{document}